\documentclass{JHEP} 
\usepackage{epsfig} 
 
\title{ Supersymmetry and the Brane World} 
 
\author{Renata Kallosh ~and ~Andrei Linde\thanks{On leave of 
absence from Stanford University until 1 September 2000}\\  
    Theory Division, CERN CH 1211 GenevA 23, Switzerland\\ 
    E-mail: \email{Renata.Kallosh@cern.ch and Andrei.Linde@cern.ch}} 
\received{\today}       
 \preprint{CERN-TH/2000-012\\ \hepth{0001071}\\ January 11, 2000} 

\abstract{ We investigate the possibility of gravity localization  on the  
brane in the context of supersymmetric theories. To realize this scenario  
one needs to find a theory with the supersymmetric flow stable in IR at  
two critical points, one with positive and the other with   negative  
values of the superpotential.   We perform a general study of the  
supersymmetric flow equations of gauged massless supergravity interacting  
with arbitrary number of vector multiplets and demonstrate that  
localization of gravity does not occur. The same conclusion remains true  
when tensor multiplets are included. We analyze all recent attempts to  
find a BPS brane-world and conclude that localization of gravity on the brane in supersymmetric theories remains a challenging but unsolved problem.}  
 

\begin{document} 
 
\section{Introduction} 
One of the most interesting recent trends   in particle physics and  
cosmology is the investigation of the possibility that we live on a  
4-dimensional brane in a higher-dimensional universe, see e.g. \cite{savas,RS}. This is a very exciting new development, and the number of new papers on  
this subject grows rapidly.  Some of these papers investigate  
phenomenological consequences of the new paradigm, some study possible  
changes in cosmology.  However, in addition to investigating the new  
phenomenology it would be  desirable to find a compelling theoretical  
realization of the new set of ideas. Here the situation remains somewhat  
controversial.  
 
In this paper we will discuss a very interesting  
possibility, discovered by Randall and Sundrum (RS) \cite{RS}. They have  
found that if two domains of 5-dimensional anti-de Sitter space with the  
same (negative) value of vacuum energy (cosmological constant) are  
divided by a thin (delta-functional) domain wall, then under certain  
conditions the metric for such a configuration can be represented as  
\begin{equation} 
ds^2= e^{2A(r)} dx^\mu dx^\nu \eta_{\mu\nu} + dr^2  \ . \label{metric1}  
\end{equation} 
Here $A(r) = -k|r|$, $k > 0$, $r$ is the coordinate corresponding to the  
fifth dimension. An amazing property of this solution is that, because of the  
exponentially rapid decrease of the factor $e^{2A(r)} \sim e^{-k|r|}$  
away from the domain wall,   gravity in a certain sense becomes localized  
near the brane  \cite{RS}. Instead of the 5d Newton law $F \sim 1/R^3$,  
where $R$ is the distance along the brane, one finds the usual 4d law $F  
\sim 1/R^2$. Therefore, if one can ensure confinement of other fields on  
the brane, which is difficult, but perhaps not as difficult as confining  
gravity, one may obtain higher-dimensional space, which   effectively  
looks like 4d space without the Kaluza-Klein compactification.  
 
However, this scenario was based on the assumption of the existence of  
delta-functional domain walls with specific properties. It would be nice  
to make a step from phenomenology and obtain the domain wall  
configuration described in  \cite{RS} as a classical solution of some  
supersymmetric theory.  Very soon after this goal was formulated, several  
authors claimed that they have indeed obtained a supersymmetric  
realization of the RS scenario. Then some of them  withdrew their  
statements, whereas many others continued making this claim.  Some of the  
authors did not notice that they obtained solutions with   $A(r) =   
+k|r|$, growing at large  $r$, which does not lead to localization of  
gravity on the brane. Some others obtained the desirable regime $A(r) =  
-k|r|$ because of a sign error in their equations. Several authors used  
functions $W$, which they called `superpotentials', but they have chosen  
functions which cannot appear in supergravity. As a result,   the  
situation became rather confusing; the standard lore is that there are  
many different realizations of the RS scenario  in supersymmetric theories,  
see e.g. a discussion of this issue in \cite{Csaki}.  The purpose of our 
paper is to clarify the world-brane--supersymmetry relation, in a large  
class of supersymmetric theories.

The simplest supersymmetric theory in 5d space with AdS vacuum is $N=2 $ 
$U(1)$ gauged supergravity \cite{GST}. The critical points of this theory  
have been studied in \cite{CKRRSW}.  It can be naturally formulated in  
AdS, so one would expect that this theory is the best candidate for  
implementing the RS scenario.  To investigate this issue it was necessary to  
find  at least two different critical points (different AdS vacua) with equal negative values of vacuum  
energy, to verify their stability (which involves investigation of the  
sign of kinetic terms of vector and scalar fields) and to find domain  
wall solutions interpolating between two different stable AdS vacua.  
 
The first important step in this direction was made by Behrndt and  
Cveti\v{c} \cite{BC}. They  found two different AdS vacuum states 
where the scalar fields have correct kinetic terms, and obtained an 
interpolating domain wall solution. However, it was not clear whether the  
sign of the kinetic terms of the vector fields is correct. More importantly,  
vacuum energies (cosmological constants) of the two AdS solutions  
obtained in \cite{BC} were different, so they were not suitable for the RS  
scenario describing a wall surrounded by two AdS spaces with equal values of vacuum energy.  
Originally, Behrndt and Cveti\v{c}  claimed that their domain wall  
solution  has properties similar to those of the RS domain wall, with $A(r) \sim - k|r|$ at large $|r|$, but later they found that this was not the case: the solution was singular at $r=0$, and $A(r)$ was growing as $+ k|r|$ at large $|r|$.
 
Soon after that, in our paper with Shmakova \cite{KLS} we  found a  
model that 
admits a family of different AdS spaces with equal values of  
the vacuum energy and with proper signs of kinetic energy for the scalar and  
vector fields. 
At first glance, it seemed that 
this provided a proper  
setting for the realization of the RS scenario  in supersymmetric theories. Indeed, we found a domain  
wall configuration  separating two different AdS spaces with equal vacuum  
energies. This configuration has  the  
 metric   (\ref{metric1}). However, 
instead of $A(r)=  -k|r| $, which 
 is necessary for the localization of gravity on the wall, we have found 
 that $A(r) \sim + k|r|$ at large $|r|$. At small $|r|$ the function 
 $A(r)$ and the curvature tensor are singular,  just as in the model of \cite{BC}.   Instead of localization of 
gravity near $r = 0$ on a smooth domain wall, there  
is a naked singularity at  $r \to 0$.  We explained in \cite{KLS} that this is a generic result which should be valid in any version of $N=2 $ 
$U(1)$ gauged massless supergravity with one moduli.
It was  also shown in \cite{KLS} that the desirable regime $A(r)=  -k|r| $ is  
impossible not only for  supersymmetric (BPS) configurations, but for any other domain wall  
solutions that may exist in this theory.

In this paper we will study $N=2$ $U(1)$ gauged massless supergravity with  
many moduli. To understand the possibilities to find the brane-world in  
such theories it is sufficient to study the behavior of the system near  
the critical points with the help of the supersymmetric flow equations.  
 
This analysis will explain  the main results obtained in   \cite{KLS}, and 
generalize them  for the theories with an arbitrary number of moduli. We  
will  
 discuss the known theories   with  hypermultiplets and  tensor multiplets 
included. We will also analyze the recent results obtained  in other 5d  
supergravity  theories  and will show that none of the solutions that  
have been obtained so far lead to localization of gravity on a brane. All  
known examples of  thick domain wall solutions with decreasing warp  
factor \cite{ST,CG,gubser,gremm} have been obtained by introducing   
non-supersymmetric `superpotentials', which cannot appear in the framework  
of a supersymmetric theory.

We do hope that it is not  really necessary to make a choice between the RS  
scenario and supersymmetry. However,  we were unable so far to find any  
simple resolution of this problem. It remains a challenge to find a  
supersymmetric extension of the RS scenario or to prove that it does not  
exist.  
 
\ 
 
\section{BPS solutions near the critical points of massless $d=5$, $N=2 $ gauged 
  supergravity interacting with abelian vector multiplets} 
 
The most clear analysis of the situation  can be  given for  $N=2 $,  $d=5 $  
gauged supergravity  \cite{GST} interacting with an arbitrary  number of  
vector multiplets, i.e. with arbitrary number of moduli. These  theories  
have critical points where the moduli are constant and the metric is an  
AdS one.  
 
The  energy functional for static $r$-dependent configurations in these  
theories can be presented as follows  \cite{ST,gubser,entr}:  
\begin{equation} 
E= {1\over 2} \int_{-\infty}^{+\infty} dr a^4 \left \{ [ f^a_i (\phi^i)'  
\mp 3f^{ai}\partial_i  W ]^2 - 12 [  {a'\over a} \pm W]^2\right \} \pm 3  
\int_{-\infty}^{+\infty} dr {\partial \over \partial r} [ a^4 W] .  
\label{energy} \end{equation} 
Here $\partial_i  W \equiv  {\partial_i  W \over \partial \phi^i}$. We  
have used the  ansatz for the metric in the form  
\begin{equation} 
ds^2= a^2(r) dx^\mu dx^\nu \eta_{\mu\nu} + dr^2  \ , \label{metric}  
\end{equation} 
where the scale factor $a(r)$ can be represented as $e^{A(r)}$. The  
energy, apart from the surface term, depends on one complete square with  
 positive sign and another one with  negative sign. Here the  
 moduli $\phi^i$ depend on $r$ and $W[\phi]$ is a superpotential 
 defined in supergravity  \cite{GST}  via the moduli fields $h^I(\phi)$ 
constrained to a cubic surface $C_{IJK} h^I h^J h^K = 6$. The moduli live  
in a very special geometry with the metric $g_{ij}(\phi)= f_i^a f_j^b  
\eta_{ab}$ where $f_i^a$ are the vielbeins of the very special geometry.  
The critical points of the superpotential were studied in  \cite{CKRRSW},  
where it was also shown that they are analogous to the critical points of  
the central charge, defining the black hole entropy. The supersymmetric  
flow equations, which follow from this form of the energy functional, are  
given by  
\begin{equation}\label{qq} 
(\phi^i)'  =  \pm 3 g^{ij} \partial_j  W \ , \qquad \qquad   {a'\over a}  
= \mp W \ .  
\end{equation} 
The warp factor of the metric  decreases at large  positive $r$ under  the
condition that  $H_r\equiv {a'\over a}$ is negative. Note that   $H_r$ is  
the space analog of the Hubble constant in the FRW cosmology, where $ds^2=  
-dt^2 + a^2(t) d\vec x^2$ and $H={\dot a\over a}$. When $H>0$ the FRW  
universe is expanding, when $H<0$ the FRW universe is collapsing. In our  
case,  
 the warp factor is increasing if  $H_r>0$,  and
decreasing if  $H_r<0$.

It is important to stress here that the sign of the superpotential $W$ is  
not a deciding factor for establishing the sign of $H_r$ since there are  
two different sets of equations, either  
\begin{equation}\label{I} 
i) \hskip 2 cm (\phi^i)'  =  + 3 g^{ij} \partial_j  W \ , \qquad \qquad  
{a'\over a} = - W \ .  
\end{equation} 
or  
\begin{equation}\label{II} 
ii) \hskip 2 cm  (\phi^i)'  =  - 3 g^{ij} \partial_j  W\ , \qquad \qquad  
{a'\over a} = + W \ .  
\end{equation} 
We are interested in critical points of the supersymmetric flow equations  
where all moduli   take some finite fixed values $\phi^i_{*}$  
\begin{equation}\label{cr} 
  (\phi_{cr}^i)'  = 0 \qquad  \Rightarrow 
  \left( \partial_i  W \right) _{cr}=0\ . 
\end{equation} 
As a result,  the superpotential $W$ acquires  some non-vanishing  
constant value defining the absolute value of the `Hubble constant' $H_r$  
at the critical point where the moduli are fixed.  
\begin{equation}\label{cr2} 
 H_r (\phi^i_{*})=  \left({a'\over a}\right)_{{\phi^i_{*}}} 
  = \pm  W  (\phi^i_{*}) = {\rm const}\ . 
\end{equation} 
We have to find out whether it is possible under some condition to find a  
solution where  $ {a'\over a} $ is 
negative at the fixed points of the moduli .  
 
It is useful to rewrite the supersymmetric flow equations (\ref{qq})  
using the chain rule $a {\partial \over \partial a} \phi^i =  a {\partial  
r \over  
\partial a} {\partial \phi^i\over 
\partial r}= {a \over a'} (\phi^i)'$. It 
follows that for any choice of equations i) or ii) we get  
\begin{equation}\label{beta} 
a {\partial \over \partial a} \phi^i =  - 3 g^{ij}{\partial_j W \over  
W}\equiv \beta^i(\phi) \ .  
\end{equation} 
This equation defines the beta functions for the supersymmetric flow  
equations reinterpreted as renormalization group equations. At the  
critical point $\phi^i_{*}$ the $\beta$-function   vanishes, $\beta  
^i(\phi_{*})=0$. This form of equations cannot be used near the vanishing  
values of the superpotential $W$. However we will use these equations  
near the critical points where the moduli are fixed at finite values  
$\phi^i_{*}$ and $W (\phi_{*})\neq 0$. To analyze the critical point we  
need to find out whether ${\partial \beta\over \partial \phi}$ is  
positive or negative near $\phi_{*}$.  
 
The universality of the critical point in this class of theories follows  
from the very special geometry in the moduli space, which allows the  
calculation of the $\beta$-function near the critical point. The basic  
relation following from the  very special geometry at the critical  
point\footnote{The analogous equation was for the first time derived in  
$d=4$ supergravity in  \cite{GFK}, which has allowed us to prove that the  
entropy of the supersymmetric black holes is a minimum of the BPS mass.}  
was derived in  \cite{CKRRSW}  
\begin{equation} 
  (\partial_i \partial_j W)_{cr} = {2\over 3}g_{ij} W_{cr} \ . 
\label{W''}  
\end{equation} 
We find out that near the critical point  
\begin{equation} 
 {\partial 
\beta^i \over \partial \phi^j}(\phi_{*})= -2 \delta^i_j \ .  
 \label{delta} 
\end{equation} 
 This means that near the critical point $\phi^i = 
\phi_{*}^i + \delta \phi^i$ the behavior of the moduli is  
\begin{equation}\label{beta2} 
a {\partial \over \partial a} \phi^i = (\phi^j - \phi^j_{*}){\partial  
\beta^j \over \partial \phi^i}  = -2 (\phi^i - \phi^i_{*}) \ .  
\end{equation} 
Therefore if the system starts at some values of moduli $\phi$  below the  
fixed value $\phi_{*}$, it will be driven to the larger  values of moduli  
towards the fixed point. If it starts at values of moduli $\phi$ above  
$\phi_{*}$, it will be driven back to smaller values of moduli till it  
will reach the fixed value. We can also solve this equation  near the  
fixed moduli $\phi^i_{*}$ (which are $a$-independent) : 
\begin{equation}\label{UV} 
 \phi^i (a)  = \phi^i_{*} + c^i a^{-2} \ . 
\end{equation} 
Here $c^i$ are some arbitrary, undefined constants.  This corresponds to  
a UV fixed point in quantum field theory; at large values of the scale  
parameter $a$, when $a\rightarrow +\infty$, the system is driven towards  
the fixed point. It follows that the stable critical point of moduli  
requires that {\it the scale factor of the metric in the direction  
towards the fixed point grows}.  
 
As we see, Eq. (\ref{UV}) is completely universal, does not depend on  
the number of moduli, choice of the cubic surface, etc. It also shows that the  
warp factor at the critical points, where the moduli are fixed, is always  
increasing as the `Hubble constant" $H_r$ is positive.  We have thus 
established   that in   $d=5 $, $N=2 $ $U(1)$-gauged massless supergravities  
\cite{GST} with an arbitrary number of vector multiplets and any choice of  
the  
 cubic surface,  the `Hubble constant' $H_r$ is {\it
 always positive} at the supersymmetric critical 
 points,  where all scalars are fixed,   and the warp factor is therefore {\it always increasing} at $|r| \rightarrow + \infty$: 
\begin{equation}\label{poit} 
 H_r (\phi_{*})=  \left({a'\over a}\right)_{\phi_{*}} = | W_{cr}| > 0\ . 
\end{equation} 
This is a prediction that, in all particular cases, if one starts solving  
the system of equations (\ref{qq}), for each  sign of $W$, the consistent  
solution for $r >0$ will be only possible in  case i) for $W<0$ and in  
case ii) for $W>0$. In one moduli case where the relevant solutions were  
presented in  
 \cite{BC} and  \cite{KLS} this was indeed the case: it was impossible to get 
the decreasing warp factor in BPS domain walls of massless gauged  $d=5 $ $N=2 $  
supergravity.  
 
Here we have shown (see also \cite{KLS}) that the deep reason for this is related to the fact 
that the second derivative of the superpotential must have the same sign  
as the superpotential at the fixed point of moduli, as shown in eq.  
(\ref{W''}). This leads to the UV fixed point universally for this class  
of theories.  
 
\section{Multivalued   Nature of the Superpotential in $N=2 $ supergravity} 
 
The investigation performed in the previous section is sufficient to show,  
even without a detailed study of the structure of domain walls in  $d=5 $,  
$N=2$ supergravity, that BPS states with gravity localization do not appear  
in this theory. However, it is still interesting to find out exactly what  
kind of vacua and interpolating solutions are possible in this theory.

The existence of the disconnected branches of the moduli space (different  
AdS vacua)  in  $d=5 $, $N=2 $ supergravity interacting with vector multiplets  
was discovered in  
 \cite{GST1} and studied more recently in  \cite{BC}, \cite{KLS} and 
 \cite{entr}. 
The existence of the disconnected branches of the moduli space leads to a  
possibility to use two different branches of the moduli space and  
therefore two different superpotentials on each side of the wall. This  
property allows in principle  a solution to be found for the scalars  
interpolating between two branches of the moduli space. However,  the  
space-time metric has naked singularities  and the warp factor  always  
increases away from the wall.  
 
Here we will explain first  of all  why different branches of moduli  
space are quite natural in  $d=5 $, $N=2 $ supergravity interacting with vector  
multiplets. The `kinky supergravity' of G\"{u}naydin, Sierra and Townsend  
\cite{GST1} has the following set up. The theory is defined by a cubic form  
\begin{equation} 
  {1\over 6} C_{IJK} h^I h^J h^K      = F \ , \qquad  F=1 \ , 
\label{cubic}  
\end{equation} 
 and a linear form 
\begin{equation} 
  W= h^I V_I \ . 
\label{super}  
\end{equation} 
The constants $C_{IJK}$ define the cubic surface and the constants $V_I$  
define the superpotential. The independent scalar fields $\phi^i$  are  
coordinates on the cubic hypersurface $F$. The restrictions on $C_{IJK}$  
is such that the metric defining the coupling of vector fields $a_{IJ}  
\sim -  
\partial_{I}\partial_J F$ for $F>0$ and the metric, induced on the surface, $g_{ij}= 
a_{IJ} h^I_{,i} h^J_{,j}$   are positive definite. There exists one point  
at the surface where $a_{IJ}$ is simply an Euclidean metric  
$\delta_{IJ}$. This is a so called `basepoint'. The existence of the  
`basepoint'  has allowed the authors of Ref.  \cite{GST} to prove that in the one-modulus case the following constants can be chosen:
\begin{equation} 
  C_{000}= 1\ , \qquad C_{011}  =- {1\over 2}\ , \qquad C_{001} =0 \ ,\qquad 
  C_{111}=C \ . 
\label{C}  
\end{equation} 
The most general case of arbitrary $C_{IJK}$ for $I=1,2$ can be reduced  
to this one. The cubic polynomial takes the following form (for $\phi=  
{h^0 \over h^1}$)  
\begin{equation} 
  F \sim (h^1)^3 \left(\phi^3 - {3\over 2} \phi + C\right) \ . 
\label{polyn}  
\end{equation} 
The discriminant of the cubic polynomial  
\begin{equation} 
  \Delta = C^2 - {1\over 2} 
\label{delta2}  
\end{equation} 
can be positive, zero or negative. In these three cases one has 1,  2 or 3  
branches of the moduli space,  respectively, i.e.  the curves $F=1$ have  
1, 2 or 3 branches. In this form the metric of the moduli space is known  
only at the `basepoint'. Thus, {\it a priori} one may have expected that  
some of the branches can be excluded if the metric is not positive there.  
 
In more recent studies of the branches of the moduli space related to  
domain walls a somewhat different set up was taken. The main emphasis was  
to find the disconnected branches of the moduli space such that the  
scalar and the vector metric are positive-definite; between branches the  
metric may be  infinite, however. The expression for the moduli space  
metric is \cite{entr}  
\begin{equation} 
g_{\phi\phi} = \alpha \left[(C_{100}^2- C_{110} C_{000})\phi^2 -(C_{111}  
C_{000} - C_{110} C_{001}) \phi + C_{110} ^2 - C_{000} C_{100}\right] \ ,  
 \label{modulimetric} 
\end{equation} 
where $\alpha >0$. For the case (\ref{C}) we get  
\begin{equation} 
g_{\phi\phi} = {1\over 2}  \alpha (\phi^2 - 2 C  \phi + {1\over 2}) \ .  
 \label{modulimetric2} 
\end{equation} 
The condition that the metric is everywhere positive is that  
\begin{equation} 
  \Delta = C^2 - {1\over 2}<0 \ . 
\label{deltaneg}  
\end{equation} 
When we combine the information from the two approaches, in \cite{GST1}  
and in \cite{KLS,entr}, we find that the case with 3 branches of the  
moduli space defined in \cite{GST1} automatically leads to the positive   
metric, whereas in cases when $\Delta >0$ or $\Delta =0$ there are parts  
of the moduli space where the metric is not positive. To have positive  
and negative  superpotentials in different branches of the moduli space  
turns out to be a necessary condition for the positivity of the vector  
space metric  \cite{KLS}. Thus in  $d=5 $, $N=2 $ supergravity interacting with  
vector multiplets with the same values of $C_{IJK}$ and $V_I$ it is  
possible to find   two distinct stable AdS critical points.

The stability of different AdS vacua is a necessary but not a sufficient 
condition for the realization of the RS scenario. One must find    
values of the parameters $C_{IJK}$ and $V_I$ that allow the existence of  
different stable  AdS vacua with equal vacuum energy density. This  
problem is rather non-trivial, but fortunately one can find several  
continuous families of parameters that satisfy this condition  
\cite{KLS}. Once this problem is solved, one may try to find an  
interpolating domain wall solution separating two different stable AdS  
vacua with equal values of the vacuum energy.  
 
One such solution of Eqs. (\ref{II}) was found in \cite{KLS}; it is  
represented here in Fig.\ref{fig1}. As we see, the scalar field grows  
from its negative critical value $\phi = -0.2$ at $r \to -\infty$ to a  
positive value $ \phi = 1$ at $r \to +\infty$.  
 The superpotential discontinuously changes its sign from negative to positive at $r = 0$. 
At large $r$ the function $A(r)$ grows as $|r|$ rather than decreases as  
$-|r|$. Thus, just as we expected, there is no gravity localization in  
this scenario.

Even though the solution for the scalar field $\phi$ smoothly  
interpolates between the two attractor solutions, the function $A(r)$ is  
singular. It behaves as $\log |r|$ at $|r| \to 0$. Metric near the domain  
wall is given by  
\begin{equation} 
ds^2= r^2 dx^\mu dx^\nu \eta_{\mu\nu} + dr^2 \ . \label{metric2}  
\end{equation} 
This implies the existence of a naked singularity at $r=0$, which  
separates the universe into two parts corresponding to the two different  
attractors.  
 
\FIGURE{\epsfig{file=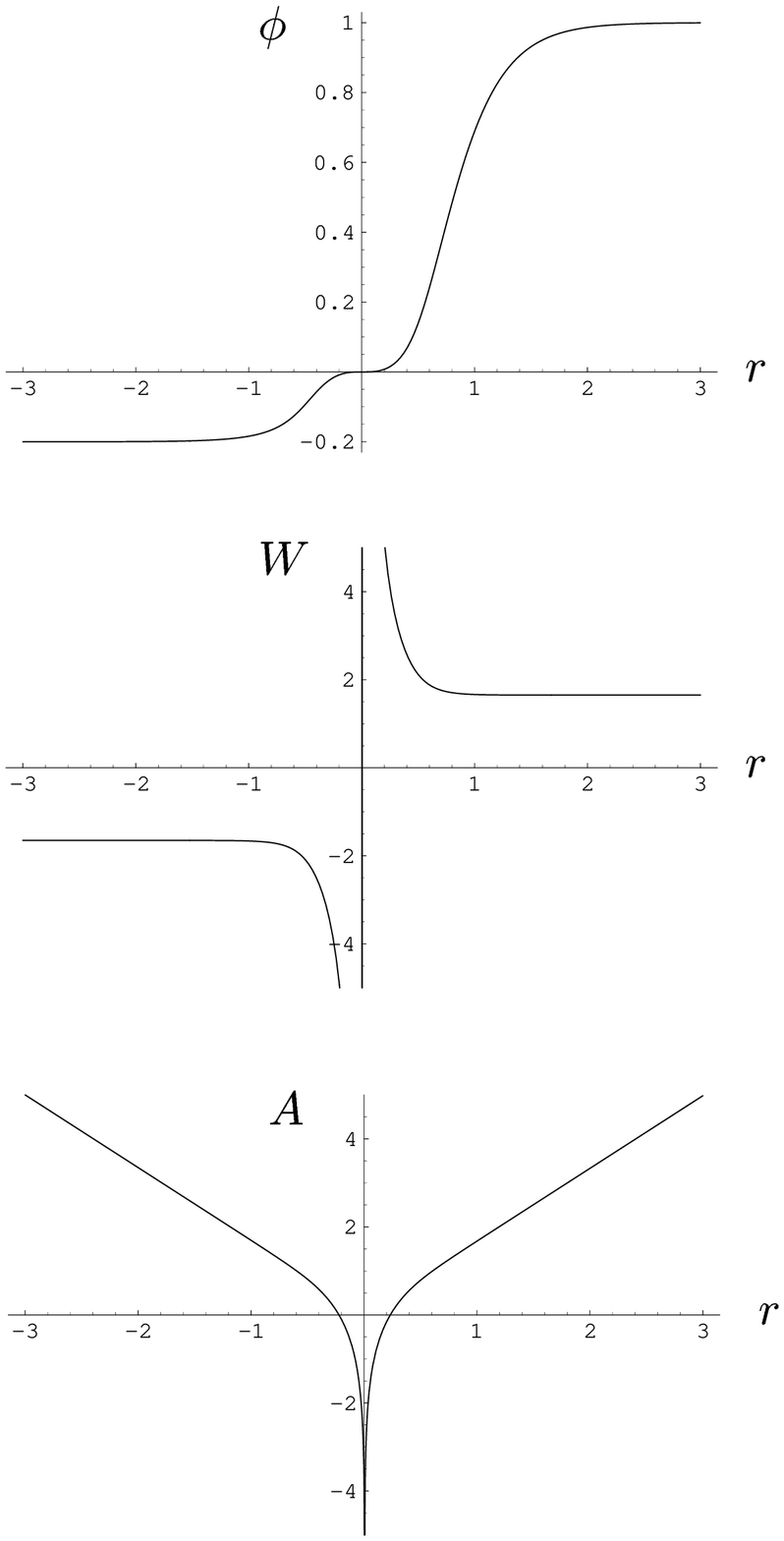,width=10cm}  
        \caption{A solution for the scalar field $\phi$ 
        interpolating between 
        two different vacua with equal values of $|W|$ \cite{KLS}. 
        Note that $\phi(r)$ is non-singular on the wall because 
        of the vanishing of $g^{-1}$ at $\phi = 0$, whereas $W$ 
        and $A$ are singular: $W \sim r^{-1}$ and $A(r) \sim \log |r|$ at $|r| \to 0$. At 
large $r$ the function  
        $A(r)$ grows as $|r|$ rather than decreases as $-|r|$. 
         This is a general property of interpolating 
         solutions 
          in our class of models.} 
    \label{fig1}}

In the same theory there is also a second BPS solution, which corresponds  
to the different choice of sign of the pair of equations, as in  Eq.  
(\ref{I}).  The solution is shown in Fig. \ref{fig3}. This configuration  
has a negative superpotential at large positive $r$ and decreasing field  
$\phi$, but both solutions have the same warp factor. This illustrates  
the statement made in the previous section that the flow equation and the  
resulting geometry of BPS states does not depend on the choice between  
the two equations (\ref{I}) and  (\ref{II}).

\FIGURE{\epsfig{file=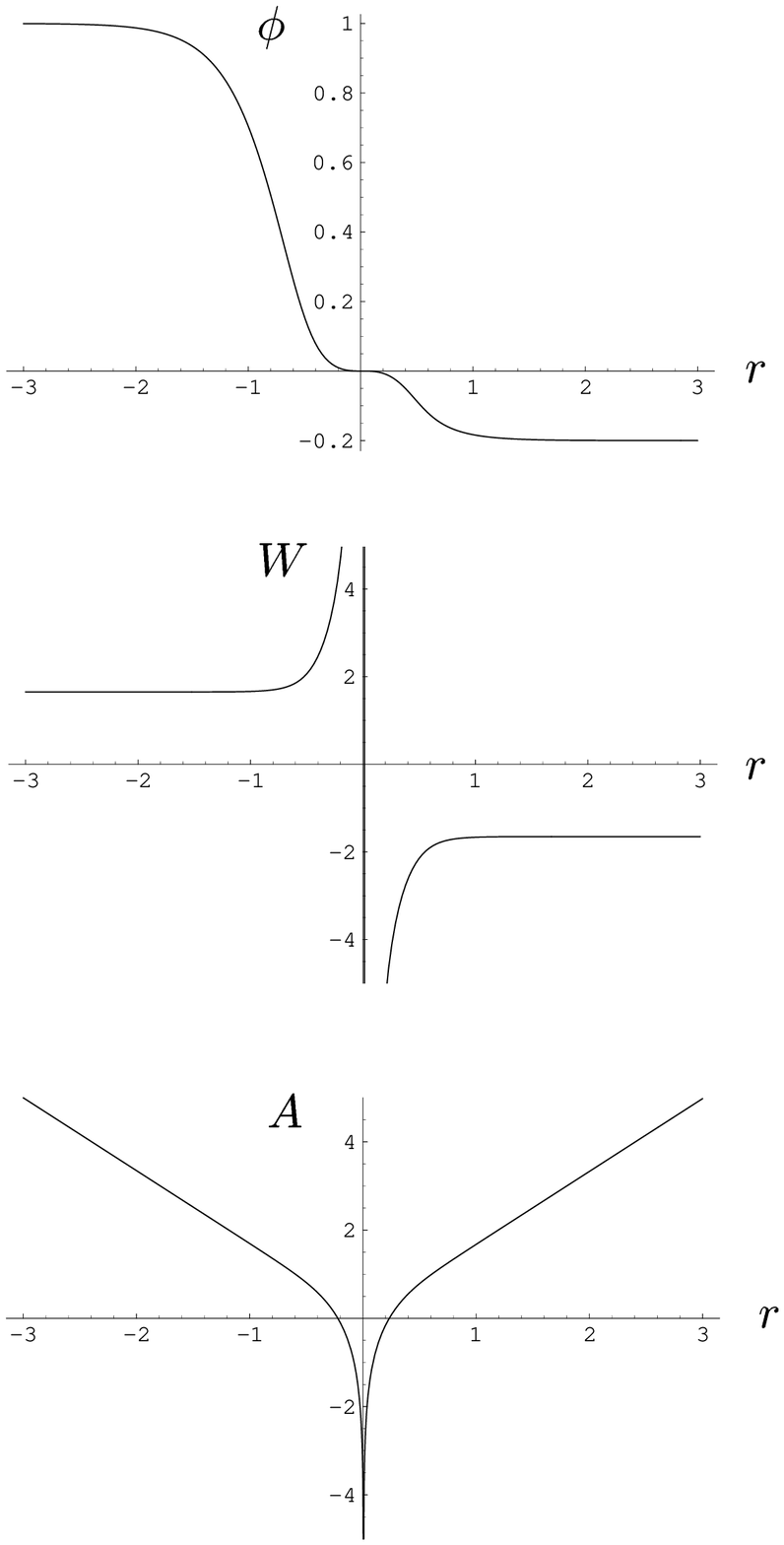,width=10cm}  
        \caption{This solution  differs from the one 
        in Fig.1 by the sign of the superpotential. Therefore the scalar field 
        interpolates between $1$ on the left and $-0.2$ on the right whereas in Fig. 1 it interpolates between $-0.2$ on the left and $1$ on the right. However, the warp  
        factor increases at large $|r|$ in both cases.} 
    \label{fig3}} 
 
We conclude that one can find more than one stable AdS critical point in  
$N=2$  $d=5 $ supergravity. However, this does not lead to localization of  
gravity on the domain wall separating two different AdS vacua.  
 
\section{Non-BPS solutions near the critical points of massless\\ 
 gauged supergravity} 
Until now  we were looking only for supersymmetric  solutions, and found  
that they do not have the desirable behavior $a \rightarrow 0$ at the  
points where scalars are stabilized (e.g. at $|r|\rightarrow  \infty$).  
One may wonder whether one  
 can find more general, non-supersymmetric  solutions with the   
  asymptotic $a \rightarrow 0$.  The answer to this question is also negative. 
  Indeed, 
  the relevant 
   equation of motion for the  scalars in the background 
   metric is 
\begin{equation}\label{deviation} 
 (\phi^i)'' + 4 H_r  (\phi^i)' + g^{ij} g_{jl, k}(\phi^k)' (\phi^l) ' - 6 
  g^{ij} \partial_j V=0\ , 
\end{equation} 
where at the critical points $ V_{,ij}$ is negative-definite. The  
potential is defined as $V= - 6 (W^2 - (3/4)g^{ij}W_i W_j)$. For all  
massless  $d=5 $, $N=2 $ gauged supergravities under discussion the second  
derivative of the potential is proportional to the potential, that is  
negative at the critical point:  
\begin{equation} 
  (\partial_i \partial_j V)_{cr} = {2\over 3}g_{ij} V_{cr} \ . 
\label{V''}  
\end{equation} 
Let us assume that the solution of this equation asymptotically  
approaches an attractor point $\phi_{*}$ at large $r > 0$, so that  
$g_{ij}(\phi_{*})$ and $g_{ij,k}(\phi_{*}) $ become constant,  and  
$(\phi^i)'$ gradually vanishes at large $r$. We will assume that $H_r$ is  
negative near the critical point since we are looking for solutions with  
decreasing warp factor. Then the deviation $\delta \phi^i$ of the field  
$\phi^i$  
  from its asymptotic value $\phi_{*} ^i$ at large positive $r$ satisfies the 
  following equation: 
\begin{equation}\label{deviation2} 
( \delta \phi^i)'' - 4 |H_r| (\delta \phi^i)' = - 4 | V|\delta\phi^i ~.  
\end{equation} 
Thus for each scalar field  we have the same  equation as for a harmonic oscillator with  
a negative friction  
 term $-  |A'| \delta\phi'$. Solutions of this equation describe oscillations 
 of $\delta\phi$ with the
 amplitude blowing up at large $r>0$, which contradicts our 
 assumptions. 
Let us explain this argument in a more detailed way. We are looking for  
asymptotic solutions of this equation at large $r$. In this limit all  
parameters of this equation take some constant values: $2|H_r| = C_1$, $4  
| V| = C_2$, where $C_i > 0$. Then Eq. (\ref{deviation2}) reads:  
\begin{equation}\label{deviation2a} 
 (\delta \phi^i)'' -2 C_1 ( \delta \phi^i)' + C_2  \delta \phi^i = 0 \ . 
\end{equation} 
Solutions of this equation can be represented as $ \delta \phi^i = e^{i  
\omega r}$, where $\omega$ may take complex values. Then this equation  
implies that  
\begin{equation}\label{deviation2b} 
\omega^2 +i 2C_1 \omega    - C_2   = 0 \ ,  
\end{equation} 
which yields  
\begin{equation}\label{deviation2c} 
\omega =  -i(C_1 \pm \sqrt{C_1^2-C_2}) \ ,  
\end{equation} 
and  
\begin{equation}\label{deviation2d} 
\delta \phi^i = e^{i\omega r} = \exp\left[(C_1 \pm  
\sqrt{C_1^2-C_2})\,r\right] \ .  
\end{equation} 
Thus at large $r$ Eq. (\ref{deviation2}) has two independent solutions.  
Both solutions grow exponentially in the limit $r\to \infty$. This means  
that our assumption that the solution can asymptotically approach a  
constant value is incompatible with the condition that $H_r  
<0$. In this proof it was essential also that  at the critical points $ 
V_{,ij}$ is negative-definite (which means that the curvature of the  
effective potential is negative). Indeed, for $ V_{,ij}  
> 0$ one would have $C_2 
<0$, and one of the two solutions given in Eq. (\ref{deviation2d}) would 
exponentially decrease at infinity, which is the required regime. But  
this regime is impossible in massless $U(1)$-gauged supergravity  where $  
V_{,ij}$ is always negative near the attractor.

\section{Solutions of other  supergravity theories with AdS critical points} 
 
Here we will give a  short overview of the possibilities.  
 
\ 
 
1. We start with a comment on  $d=5 $,  $N=8$ gauged supergravities. For  
these theories one finds out that, in known cases of supersymmetric flow  
equations  presented in the literature \cite{Khavaev,Freed,Bakas},  the  
first order BPS-type equations have the form $H_r = -{1\over 3} W$ and  
the superpotential at all known critical points  is negative. Some of  
these critical points with maximal unbroken supersymmetry have a UV fixed  
point behavior, some have saddle points  with smaller supersymmetry  
unbroken and have a IR point behavior. However since the superpotential  
is negative at all known critical points one cannot realize the situation  
that $H_r$ is positive  at $r\rightarrow +\infty$, which would correspond  
to a decreasing warp factor away from the wall in the positive $r$  
direction.  
  No such 
theory seems to be available in the literature.  
 
\ 
 
2. A version of  5d supergravity interacting with the vector multiplets  
and the so-called universal hypermultiplet was found in \cite{kelly}. If  
the hypermultiplet is gauged, there is a contribution to the potential,  
which does not allow an AdS vacuum in this theory. If the gauging of the  
universal hypermultiplets is removed, the AdS critical points are  
possible, however, they are defined by the vector multiplets exclusively.  
The problem is reduced to the one that was studied before and there is  
no world-brane BPS walls with decreasing warp factor.  
 
\ 
 
3. The recently discovered $N=2 $,  $d=5 $ gauged supergravity with vector and  
tensor multiplets  \cite{murat} has a new type of a potential:
\begin{equation} 
  V=2  g^2 W^{\tilde a} W^{\tilde a} +g_R^2 (-P_0^2 + P^{\tilde a} P^{\tilde 
  a}) \ . 
\label{ten}  
\end{equation} 
The scalars from vector multiplets give the usual contribution, proportional  
to $g_R^2$; here $P_0$ is a superpotential and $P^{\tilde a}$ is  
proportional to the derivative of the superpotential over the moduli. The  
new potential has an additional contribution, proportional to $g^2$,  due  
to tensor multiplets, which is manifestly non-negative. The BPS form of the action consists of 3 full squares, i. e. in addition to all terms in eq. (\ref{energy}) there is a positive contribution to the energy  $2  g^2 W^{\tilde a} W^{\tilde a}$.

To understand the  
situation near the critical points in this class of theories, consider  
the supersymmetry transformations (with vanishing fermions)  
\begin{eqnarray} 
\delta \psi_{\mu}^{ i}  & = & \nabla _\mu \epsilon^i + {i\over 2\sqrt 6}  
g_R P_0 (\phi) \Gamma_\mu \delta ^{ij} \epsilon_j \ ,\nonumber\\ \delta  
\lambda^{i\tilde a} & = & -{i\over 2} f^{\tilde a}_{\tilde x} \Gamma^\mu  
({\cal D}_\mu \phi^{\tilde x}) \epsilon ^i + g W^{\tilde a} \epsilon^i  
+{1\over \sqrt{ 2}} g_R P^{\tilde a}(\phi) \delta^{ij} \epsilon_j \ .  
\label{murat}  
\end{eqnarray} 
At the critical point where the moduli are constant the unbroken  
supersymmetry requires that  
\begin{eqnarray} 
 \delta 
\lambda^{i\tilde a} =  g W^{\tilde a} \epsilon^i +{1\over \sqrt{ 2}} g_R  
P^{\tilde a}(\phi) \delta^{ij} \epsilon_j =0 \ .\label{murat1}  
\end{eqnarray}

 If the full $N=2 $ supersymmetry is unbroken at the critical point, we  
have to require that \footnote{We have learned that M.  
G\"{u}naydin has  found the same condition for the supersymmetric fixed  
points in this theory (private communication).}.  
\begin{equation} 
 g W^{\tilde a}(\phi_*)= {1\over \sqrt{ 2}} g_R P^{\tilde a}(\phi_*) =0 \ . 
\label{enhanc}  
\end{equation} 
 Without tensor multiplets near the critical point where the scalars are not fixed,  the $r$-derivative of  scalars 
is proportional to the derivative of the superpotential and only 1/2 of  
supersymmetry is unbroken. In presence of tensor multiplets we may try to relax the condition for the critical point and request that $ g W^{\tilde a}(\phi)\neq 0$. One can verify that this is not possible if any supersymmetry is unbroken.  Consider  a condition that all 3 bosonic terms in the gaugino supersymmetry transformation are not vanishing:
\begin{equation} 
 \phi' \sim  W^{\phi} \sim  P^{\phi} \ .
\label{noenhanc}  
\end{equation} 
 We have to find a projector specifying the Killing spinor. Under  
such condition the Killing spinors in addition to the usual constraint  $i\Gamma^r \epsilon^i = \pm \delta^{ij} \epsilon_j$ have to  
satisfy the following condition:  
\begin{equation} 
\epsilon ^i =\epsilon^{ij} \epsilon_j =\pm \delta^{ij} \epsilon_j \ .  
\label{killing}  
\end{equation} 
It can be verified that this is possible only if $\epsilon_i=0$, which means  
that supersymmetry is completely broken.  However if we request that
\begin{equation} 
 \phi'  \sim  P^{\phi} \ , \qquad  W^{\phi}=0
\label{1/2}  
\end{equation} 
we find the usual Killing spinor projector for 1/2 of unbroken supersymmetry which is also consistent with the gravitino transformation.  Also the gravitino transformation  
rules have integrability condition for the existence of Killing spinors; 
this tells us, as in the case of vector multiplets only, that the AdS curvature  
is defined by the value of the superpotential $P_0^2(\phi_*)$.

This brings us back to the previously studied situation with vector  
multiplets only, where we know that $H_r >0$ at positive $r$.  Therefore we conclude  
that the supersymmetric critical points of $N=2 $, $d=5 $ gauged supergravity  
with vector and tensor multiplets  \cite{murat} have the same nature as  
the ones without tensor multiplets and therefore will not support the  supersymmetric
brane-world scenario.  

The non-supersymmetric solutions in this theory require an additional investigation.
 
\ 
 
5. Dilatonic domain walls were studied by Youm \cite{youm} in the context  
of the brane-world scenario. It was found  that the warp factor in the spacetime  
metric increases as one moves away from the domain wall for all the supersymmetric dilatonic domain wall solutions obtained from the (intersecting) BPS branes in string theories through toroidal compactifications. 
     
\ 
 
6. An interesting  development was pursued recently in the framework of  
the massive gauged supergravity in  \cite{CLP}. A short summary of the  
situation is the following. The model has one AdS critical point with the fixed scalars. It has an IR behavior  
since at the critical point, in notation of  \cite{CLP}, the   
$\beta$-function    in the supersymmetric flow equations has the opposite  
sign from that of $W$ and  
\begin{equation} 
 (\partial_i \partial_j W)_{cr} = -{2\over3}W_{cr} \ , 
\label{W''1}  
\end{equation} 
and therefore one finds that $\phi = \phi_* + a^4$. Since $W$ is negative  
at this critical point, it gives a decreasing warp factor at  
$r\rightarrow -\infty$. The second AdS critical point is absent: there is  
only a run-away dilaton behavior. Therefore this solution does not lead  
to the localization of gravity.

\section{Non-supersymmetric choice of the `superpotential'} 
 
In addition to supersymmetric theories, one may consider  non-supersymmetric
theories with potentials that can be represented in a form $V= -{1\over  
3} W^2 + {C\over8}W_\phi^2$. This resembles potentials in supersymmetric 
theories with superpotential $W$, where the constant $C$ depends on the choice of the theory \cite{ST,CG,gubser,gremm}.
Then one may choose the  
`superpotential' $W$ in a way that the brane world scenario has the  
desirable solution with decreasing warp factor away from the wall in both  
directions. As follows from our analysis,   one  
should find in the examples of this kind two IR critical points with the opposite signs of the
`superpotential'. This means that on the wall the `superpotential' must  
vanish. The second derivative of the `superpotential' must be positive  
(negative) when the `superpotential' is negative (positive). This is  
indeed the property of the solutions found in \cite{gubser,gremm}. For  
example, in \cite{gremm}  
\begin{eqnarray} 
W(\phi)= 3 \sin {\sqrt{2\over 3} \phi} \ , \quad \phi(r) =\sqrt 6 \arctan  
(\tanh{r/2})\ , \quad a(r)= e^{A} = {1\over 2\cosh r} \ .  
 \label{gremm2} 
\end{eqnarray} 
At the right critical point at $r\rightarrow +\infty$, $W$ is positive  
but $W_{\phi\phi}$ is negative. At the left critical point at  
$r\rightarrow -\infty$, $W$ is negative but $W_{\phi\phi}$ is positive.  
Therefore at both critical points (with $H_r =  -  W/3$ and $\phi' =  
 W_{,\phi}/2$)  one has
\begin{equation} 
  (\partial_\phi \partial_\phi W)_{cr} = - {2\over 3} W_{cr}\ , 
\qquad  
 \Rightarrow  \qquad \phi - \phi_* \sim a \ . 
\label{crit}  
\end{equation} 
At both critical points, $a\rightarrow 0$ is a stable point where the  
scalar field  reaches a fixed value. At $r=0$ the `superpotential'  
vanishes, so that to the right from the  
wall it is positive and to the left  it is negative.  
 
The solution for the `superpotential' proposed in \cite{gubser}  has  the  
same basic  features   as the one in \cite{gremm}  
\begin{eqnarray} 
W(\phi)= 2 (\phi - {1\over 3} \phi^3) \ , \qquad \phi(r) = \tanh r\ .  
 \label{gremm3} 
\end{eqnarray} 
\begin{equation} 
  (\partial_\phi \partial_\phi W)_{cr} = - 3 W_{cr}\ , 
\qquad  
 \Rightarrow  \qquad \phi - \phi_* \sim a^{9/2} \ . 
\label{crit2}  
\end{equation} 
 We plot this solution on Fig. \ref{freed}. 
Note that the function $A(r)$ has a desirable behavior at large $|r|$.  
 
\FIGURE{\epsfig{file=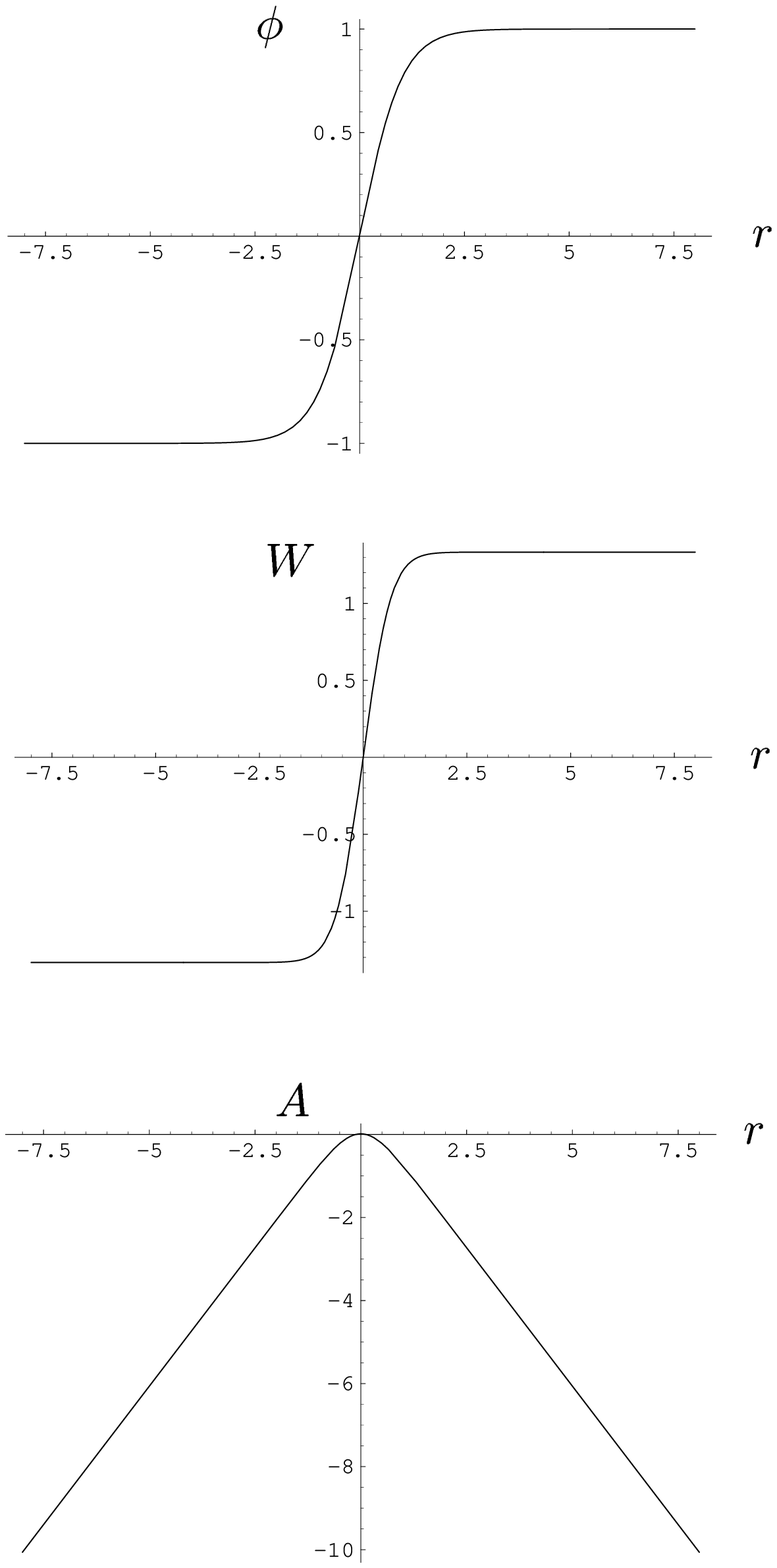,width=10cm}  
        \caption{A domain wall solution for the scalar field $\phi$ 
        in the theory with the `superpotential' $W(\phi)= 2 (\phi - {1\over 3} \phi^3)$  
 \cite{gubser}. At 
large $r$ the function $A(r)\sim -|r|$, just  as required for gravity  
localization. However, the theory with this `superpotential' is not  
supersymmetric.}  
    \label{freed}} 
     
It may be useful to compare these solutions with the BPS ones in Figs.1,  2.  
One observes that in the  non-supersymmetric  case in Fig. 3 the  
`superpotential'  vanishes at $r = 0$ and changes its sign there. Meanwhile in the  
supersymmetric case,  Fig.1 and 2, the true superpotential  
 changes   sign on the wall,  but it goes through a discontinuity. 
 
 Thus,  in those cases where the brane-world scenario can be realized 
  the `superpotential' has the following  basic features 
 (with the choice of flow equations $H_r =  -  W/3$ and $\phi' =  
 W_{,\phi}/2$):  
\begin{equation} 
  W_{r=0}=0\ , 
\end{equation} 
and also  
\begin{eqnarray} 
W_{r\rightarrow -\infty} &<& 0 \ ,   \hskip 1.7 cm  W_{r\rightarrow  
+\infty}  
>0 \ , \nonumber\\ 
 (\partial_\phi^2 W)_{r\rightarrow 
-\infty}& >& 0 \ , \qquad  (\partial_\phi^2  W)_{r\rightarrow +\infty}<0  
\ .  
 \label{nonsusyW} 
 \end{eqnarray} 
 Note that in 
both cases discussed above, at both critical points,  
$(\partial_\phi \partial_\phi W)_{cr}$ and $W_{cr}$ have opposite 
signs, which is impossible in massless gauged supergravity. 
No supersymmetric embedding have been found for such 
`superpotentials' so far. Thus the use of the word 
`superpotential' is not quite appropriate here since 
it makes an incorrect impression that the theory with the 
`superpotentials' described above is supersymmetric. 
 
The study of non-linear perturbations around such non-supersymmetric  
solutions was performed in \cite{CG}, where it was found that some `pp  
curvature' singularities appear at large $r$. Interestingly, these  
singularities at large $r$ do not appear when the proper supersymmetric  
superpotentials are used. A closely related singularity at the AdS  
horizon was discussed in \cite{CHR}, where the study of the black holes on  
domain walls was performed. These issues require further investigation.  
 
Another problem is related to quantum effects. Usually, after taking into  
account one-loop corrections,  the effective potential  in  
non-supersymmetric theories cannot be represented in the form  $V=  
-{1\over 3} W^2 + {C\over 8}W_\phi^2$. Therefore the notion of  
`superpotential' becomes irrelevant and, instead of solving first-order  
equations for BPS-type states, one should investigate solutions of the  
usual second-order Lagrange equations.

\

In conclusion we would like to point out that the analysis performed here  
shows that the brane world is not yet realized   as a BPS or non-BPS  
configuration of supersymmetric theory. We cannot exclude, however, that  
some supersymmetric theory can be found where such brane world  may exist, providing  
a consistent  alternative to compactification. As the present  
investigation shows, it may be rather non-trivial to find such a theory, if  
it exists, since the most general 5-dimensional supergravity theories  
have not been constructed yet. Since the main result depends on the  
specific sign of the beta function (in two critical points), one would  
not like to miss the existence of the correct theory. It was non-trivial   
to replace the positive $\beta$-function in QED by a negative  
$\beta$-function in non-abelian gauge theories. We hope that the analysis  
performed here will help to make a final conclusion on the compatibility  
of supersymmetry with the brane world scenario.

\

We are grateful to E. Bergshoeff, A. Brandhuber, S. Ferrara, G. Gibbons,  
M. G\"{u}naydin, J. March-Russell, R. Myers,  K. Stelle, P. Townsend, A. Van Proeyen and D. Youm for   discussions. This work was supported in part by NSF grant  
PHY-9870115.

\end{document}